# MODEL OF SOLUTIONS FOR DATA SECURITY IN CLOUD COMPUTING

Zuzana Priš áková[1] and Ivana Rábová[2]

[1]Department of Informatics, Mendel University, Brno, Czech Republic
`zuzana.priscakova@hotmail.com`
[2]Department of Informatics, Mendel University, Brno, Czech Republic
`rabova@mendelu.cz`

## ABSTRACT

*The aim of this paper is to develop a model to ensure data stored in the cloud. Model based on situations that arise in a business environment. The model also includes individual participants and their data operations. Implementation of the model is transferred using UML. The model is divided into 7 modules. Each module is apparent from the terms of data security and described specific situations when working with data. Based on this model it is possible to convert the implementation of cloud into enterprise environments with respect to data security in the firm.*

## 1. INTRODUCTION

Storing data in the cloud can be considered quite attractive form of outsourcing focused on daily data management [1]. Despite this claim, but the real responsibility for managing the data falls within the company that owns the data. With this in mind, it is important to understand some of the causes of data corruption. Such causes advise keeping the big responsibility of a cloud services, some basic best practices for the use of secure data storage to the cloud, as well as the methods and standards for monitoring the integrity of the data regardless of data storage [2]. In order to achieve higher security and redundancy of the data stored in the cloud at the same time locally.

One of the main advantages of storing data in the cloud, is unlimited access to the data, with no limitations lies in the time and place of access [3]. This property is used by firms whose occupation takes place in various remote locations. For such companies it pays to enter into cloud solutions and, therefore, that this eliminates the burden of physical storage devices, use the same computer and multiple access data in real time (real-time reporting) [3]. In this case, it is important to create storage cloud to think about the specialty store. Although there are hundreds of cloud storage, each storage site is oriented to other requirements, such as storing communication by e-mail, store employee profiles, documentation storage projects, etc. [3]. Of course, requirement may also store all types of documents.

Data integrity is critical for each data center [4]. Monitoring integrity in the cloud need to store data in the cloud. Data corruption can occur at any level of storage. Bit rot (weakening or loss of bits of data on storage media), regulator failure, damage to reduplication metadata, tape failure to advise all examples of different types of media causing damage [5]. Metadata damage may result from any of the above errors, but also are vulnerable to software bugs out rates of hardware faults [6]. Unfortunately, side effect is that the reduplication corrupted file, the block or bit affects all parts of the data associated linked to those metadata. The truth is that the data corruption can occur anywhere in the storage environment. Data can be damaged easily migrating to another platform, thus sending data to the cloud. Data storage in cloud is data centers with hardware and software that are constantly exposed to possible damage to data [7].

Despite the above minuses, is important to say that the cloud cannot be regarded as improper storage, but rather we can say that in investigating and implementing cloud strategies, there are more factors of choice, just as the cost per gigabyte stored. Data storage in the cloud offers many benefits to businesses of all sizes. Regardless of how or where the data is stored, it is imperative that the company assured that your data will be accessible in any case necessary. Therefore we used tracking data integrity and authentication.

## 2. METHODOLOGY AND RESOURCES

Data security can be defined as the confidentiality and integrity of data processed by the organization. In scenarios where the data owner has no control over the detailed architecture and management controls, such as outsourcing, assuming an increased risk of data security. We mitigate the risks by knowing the element:

- organization structure that properly values, protects and uses data, both in planning as well as the provision of services [8],
- strong and clear accountability procedures, recognizing that the data owner (organizational unit) is the best place to understand and address the risks to their information, including personal data [8],
- measures taken on the level of security of archived data, creating confidentiality, data security and sharing [7],
- establish a clear policy to be simple to understand and use [8],
- control of external parts of the business, understanding what suppliers are doing business and their control, as appropriate [5],
- provide a consistent and universal framework for safety training [9],
- clarification of the life cycle of data associated with employees of the company [5].

The safety data are often in a corporate environment uses data classification scheme. Its aim is to further highlight the necessary controls various data types that are processed businesses. The scheme of classification data is made based on legal, regulatory and business requirements that the company must follow [8]. For data security is important to mention the fact that there are general principles that could drive business to ensure data protection. However, in this time are exist norms for privacy policy. Poynter [10] sets out ten key rules to ensure the protection of data, which we can say that a more general nature:

1. information about the entity (be it an individual or firm) is attributed to each object and represent the basic general information about the entity. These data are also available to other parties, but the owner of the data is the only owner operator,
2. operator's responsibility to maintain their data,
3. data becomes information when they have significant value for the company. Usually this occurs through aggregation and context. The target in this case is to achieve a minimum of zero information loss and disabling unwanted access to information. Segregation is separated in storing data and determined jobs and systems that require information,

4. company should have a minimum of information required to perform its functions, including storage for keeping data. They should maintain the data that is available elsewhere, but in the case of routine data should be used in other data sources (not corporate), which is achieved by improving the ability to tailor their services to their customer,

5. enterprise should maintain all information about the entity together and divide them into multiple repositories. Appropriate data storage structure would be to shift data into a single entity customer record for individuals and one customer record for the company,

6. for effective information security is important to the customer and the service provider has played a role. Requires the service provider and the customer play a role. Firms should have powers to enter a secure method of exchanging data with their customers, ranging from enterprises and individuals, including the time,

7. enterprises should take into account external sources of guidance on information security, such as legislation on data protection guidelines for the financial services sector,

8. data security measures should be focused on areas of greatest risk such as data transfer. Transfer of digital data relating to the physical media should be as small as possible,

9. ongoing communication via hard copy (paper documents) should be rationalized in terms of content and frequency site with significant long-term plan their elimination,

10. computers (and soon, other removable media) should be encrypted, so if they are lost or stolen, it prevented access to any data or information.

The security features that can be inferred and therefore that the record keeping is better to have a secure area in company business, such as a data center than data stored in laptops or other devices that are located in public areas. Thin client technology, such as a browser or a terminal session, allowing access to applications, but for storing data in a data center configuration is needed, which prevents local printing store and other media [10].

If data is to be stored on computers or portable devices that come into contact with public spaces, it is necessary to encrypt data [11]. Identification, authentication and authorization controls are crucial for controlling access to information on awareness of basic information. For unstructured data types, such as e-mails, spreadsheets and word processor documents, it is advisable to use a tool DLP (Data Loss Prevention), which are likely to have a higher value [12]. DLP tools can determine what types of data are transmitted and stored in the enterprise information system, and also determine the business rules with these data, the data can be saved, printed or sent.

## 3. RESULTS AND DISCUSSION

For data security solutions, we decided to create a model by UML diagram. Use case diagram (use case diagram) was used to show integrity and confidentiality for data security. Diagram is divided into several groups framework [13]:

1. integrity and confidentiality of data-if the company relies on commercial service provider for data transmission services as a commodity item, rather than as a fully managed service, it may be more difficult to obtain the necessary assurances regarding the implementation of needed security controls for the integrity and confidentiality of transmitted data. In this diagram, we modeled a situation where a company uses encryption mechanisms for recognizing changes in data transmission and encryption systems to prevent unauthorized disclosure of information during transmission unless otherwise protected by alternative physical measures. Improvements would consist of creating a protected distribution systems,

2. input data - Information Lifecycle Management provides processes such as data classification, storage, use and disposal of rights-based regulatory and business requirements,

3. monitoring of safety - the company monitors the security controls in the information system continuously. Continuous monitoring activities include configuration management and control information system components, security impact analyzes of changes to the system, ongoing assessment of security controls and reporting status. Enterprise assesses all security controls in the information system during the initial security accreditation. The organization also provides timing control monitoring to ensure adequate coverage achieved. Terms of security controls that are critical to protecting a computer system or a small evaluation at least once a year. All other controls are assessed at least once during the accreditation of the information system. Rigorous and well executed continuous monitoring process significantly reduces the effort required for re-accreditation of the information system. Improving monitoring, we could achieve expansion and maximizing the value of ongoing assessment of security controls during the continuous monitoring process. Independent certified agent or team will review all safety checks during the deployment of the information system,

4. manager module - this section we describe more use case diagrams. The client module refers to the end user, since the module is to establish appropriate controls that should be applied to all clients who process information and mediate access to other information systems. This module refers to the whole architecture

5. accounts management- the company manages information system accounts, including the formation, activation, modification, revision, off and removing accounts. Account management includes identifying the type of account (ie, individual, group and system), establishment of conditions for group membership, and assignment-related permits. An enterprise determines the authorized users of the information system and specifies access rights. The company provides access based on need, the requirement, which is determined by assigned official duties and fulfills all the criteria for security personnel, as well as the intended use of the system. The enterprise has to have submitted applications for opening an account in the information system, as well as a summary of such applications approved. Company specifically authorizes and monitors the allocation and use of anonymous accounts and removes, disables, or otherwise secures excess bills. Manager user accounts are notified when they are terminated or transferred to the user accounts associated with another account. Supporting the management of the company accounts using available mechanism (software). The information system automatically terminates temporary and emergency accounts expired after a designated time period for each type of account. Information system automatically shuts down inactive accounts after the defined period of time. Enterprise uses automated software to audit accounts - creation, modification, disabling, and termination actions, notice to the persons concerned,

6. security risks - company develops, disseminates, regularly maintained precautions for security risks. Undertaking shall be treated formally documented risk assessment policy that addresses whether the necessary extent permitted entities, responsibilities, commitment management, coordination, and the like. The hedged risk leaving a formal documented procedures to facilitate the implementation of policies relating to risk assessment and control of risk assessment,

7. sensitive data - company assigns access to sensitive data based on user role determined using management accounts. Within this module, the risks assessed and updated, as changes in assigned security level is reflected in the risk assessment.

The following figure 1 shows the individual elements of data security, which we described in the preceding paragraphs. Also here we can see the different actors and their assignment to the module. This diagram is a rough outline of data security in a corporate environment.

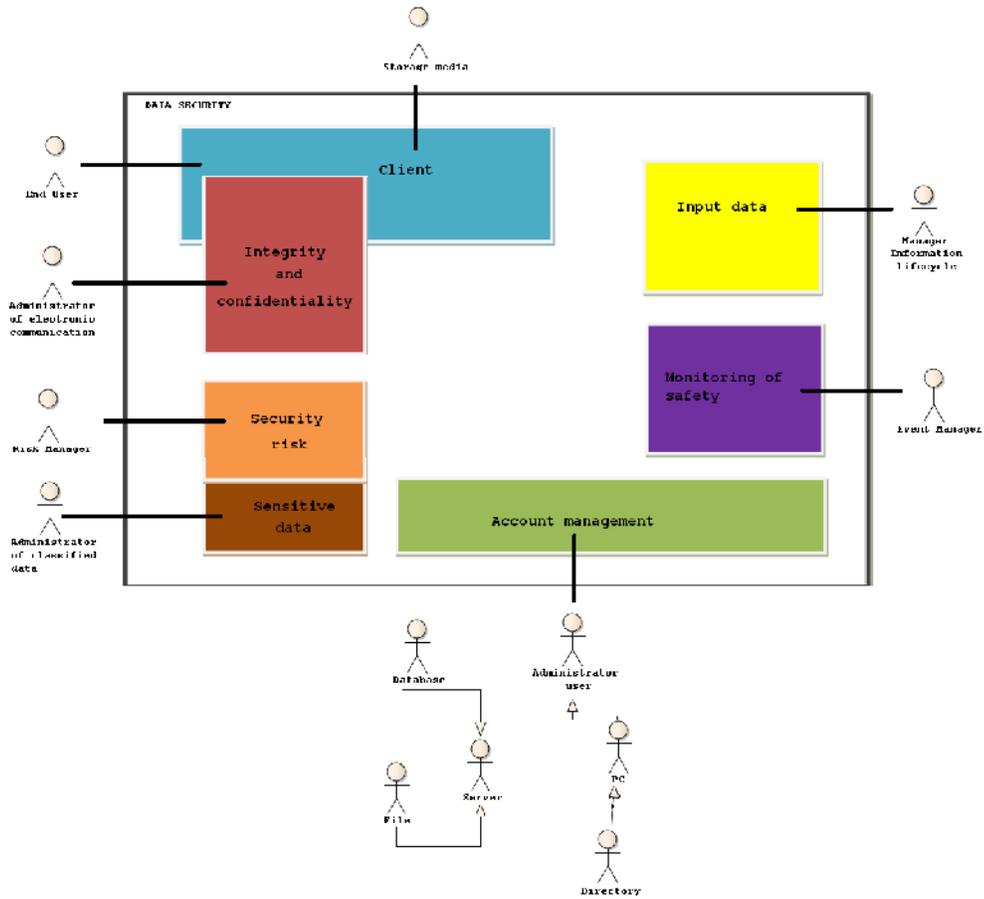

Figure 1.  Security model

We describe functionality of the model via basic algorithm. We specify the input variables using the initialization values:

```
name:='name';
passw:='password';
data:='data';
log:=false;
verify:=false;
secured:=false;
```

Variables log, verify and secured are of boolean type. We assume that name, password and data are of string type. The main procedure is called dataSecurity. This procedure is consists of procedures:

- loggedUser – for logged user via name, password,
- verification - to verify the account verification by the security of the data via name, password, data,
- privacyPolicy – setting security policy for data,
- authenticateData – authentication data,
- encryptData – encrypt data,

- authenticationCode - assignment of authentication codes,
- secureData – verification of data security.

If the value of verification is false, it means that the user does not have access to the provision of the data. The procedure dataSecurity will be concluded. Output is message that informs about data security. These procedures we are writing using UML.

```
procedure dataSecurity
begin
        procedure loggedUser(name,passw,log) ;
        if (log!=true) then procedure loggedUser(name,passw,log)
                else
                begin
                        procedure verification(name,passw,data,verify);
                        if (verify!=true) then exit
                        else
                        begin procedure privacyPolicy(data);
                                procedure authenticateData(data);
                                procedure encryptData(data);
                                procedure authenticationCode(data);
                        end;
                        procedure
                        secureData(data,authenticationCodes,secured);
                        if (secured!=true) then message('wrong secured')
                        else message('secured');
                end;
end;
```

## 2.1. Use case diagram

Diagram that illustrated in Figure 1, we specified in detail by assigning different use cases linked to the relevant actors and modules. The closer specification is shown in Figure 2.

Figure 2.  Use case diagram

In the use case diagram we show the main actors, who are involved in data security. Important actor is just an employee firm that works with data - the end user. The end user works with data through the client. He shall be required to be trained in data security system to know basic information about data security and recognized the specific requirements for sensitive information in a business context. It is important to note that the end-user to log into the information system user authentication is required, which usually consists of a username and password. These various use cases of the actors are grouped not only the client module, but also the integrity and confidentiality of data.

Other important elements are advised storage media on which data is stored. Storage media in this case, we can understand the printed document. It is important that these media are controlled, respectively, are not used at all, thereby preventing the loss of confidential data due to printing or copying to removable media. If we cannot avoid the use of storage media, it is appropriate to introduce prevention of data loss caused by printing or copying in the form of control over the transmitted port. Another important element is the encryption of sensitive information. Risk manager works with the storage media, since the compilation of the policy process to address risks permits and prevents manipulation when saving data. They also identify high-risk procedures that may occur when saving data. Event Manager is responsible for monitoring the safety of which we have already described in the definition of the individual modules. Classified Data Manager works under the model of sensitive data, which cooperates with risk managers. Its main activities include assigning security levels. Manager user accounts are divided into several groups. It is for this reason that we wanted to indicate what is responsible and what works management accounts. For a more detailed descriptions of these actors have created a common use case diagram for the database file. Business manager lifecycle information´s has been described by the module input data.

## 2.2. Sequence diagram

By accessing this diagram is a requirement of the end user to ensure created or altered data. In order to be implemented this requirement, it is necessary that the end user is logged in through the client environment. This request requires authentication by user name and password. The registration is important to verify the authorized allocation of responsibilities logged-on user. To verify the sending request framework for managing the accounts. Based on available information, then query enters a framework for the integrity and confidentiality of data, while here include the identification of security risks. Before the authentication data should be encrypted data. Encryption takes place in the module management accounts and did cooperate with the module sensitive data. Through the framework of sensitive data we determine the security level of the data that we want to encrypt, and then we can allow data security and encrypt data. The encoded data is sent to the input data framework, where the authentication codes assigned for maintaining the integrity and confidentiality of data. Finally, the end user receives the return message about data security. In the diagram entity are server and computer (Figure 3).

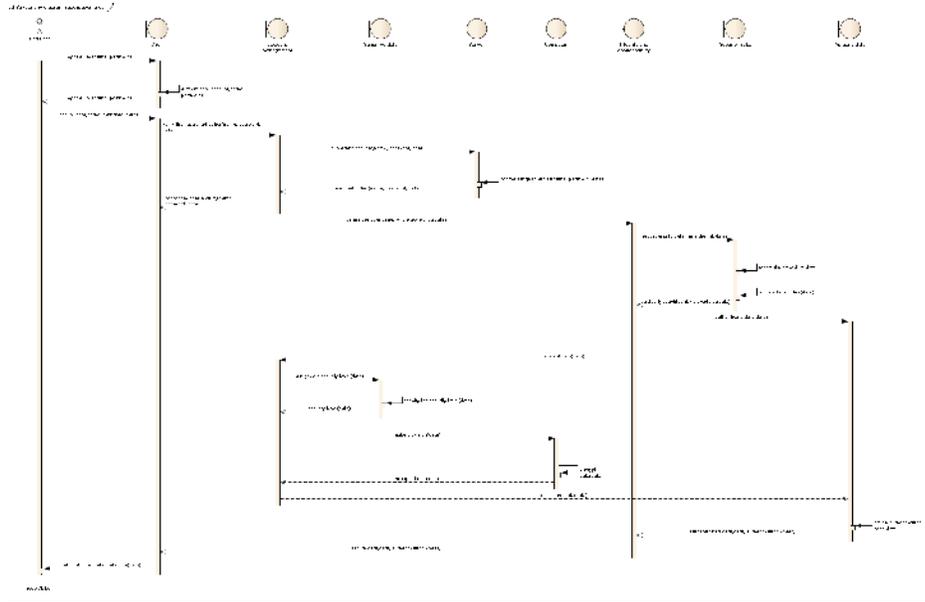

Figure 3.  Sequence diagram

## 2.3. Activity diagram

To express our individual activities and their progress in crucial situations, we decided to create activity diagram (Figure 4), and various activities fit in the query sequence diagram.

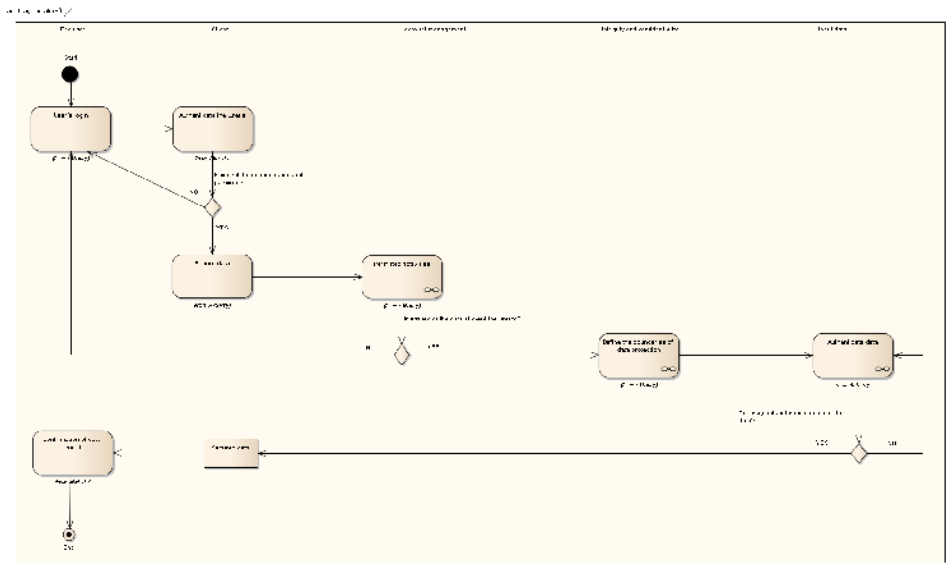

Figure 4.  Activity diagram

Since the whole process is quite complex, most complicated activities (authorized activities, to draw the line data protection, authenticate data) are denounced as structured. The diagram activities were treated authenticate the user, and when you fail we ask you to enter your credentials. Re-requesting data may occur if the current user is authorized activities with the given data.

To determine the permitted activity logged in user, you need to be verified. Individual authorizations are stored on servers in the module management accounts. If the user does not have any assigned duties, shall terminate automatically given activity. If they are given certain responsibilities, determine the assigned duties and on the basis of them enters agreement/ disagreement with the provision of data required by the user.

After identifying the authorized activities and permit data security should be determined by boundary data protection. This structured activity is directly related to the provision of risk and its various activities. Output is intended risky procedure. This activity is important to mention that also includes measures to determine risk policy, thereby achieving a more secure type of data processing in the future and call for data security.

Authentication consists of two activities. First we need to encrypt the data, and then we can assign the encrypted data authentication codes. Since data encryption is complicated activity, we modeled it as a structured activity.

The encrypted data is necessary to determine the level of data security. In this activity, we distinguish very sensitive data from less sensitive data and assign them based on their content security. We can enable data encryption after assignments. Output in structures activities are therefore encrypted data.
Once you encrypt, data are assigned authentication codes, which shall terminate ensure the integrity and confidentiality of data. As already mentioned, the result of the security data is also displayed to the end user, and thus ends the process of data security. In the process, it is necessary to log off a user of an information system, since we assume that the end user will still work with other data and will again require data security.

## 4. CONCLUSIONS

Cloud computing is a modern innovative technology of solution of a problem with data storage, data processing, company infrastructure building and so on [1]. Many companies worry over the changes by the implementation of this solution because these changes could have a negative impact on the company [1], or, in the case of establishment of new companies, this worry results from an unfamiliar environment. Data accessibility, integrity and security belong among basic problems of cloud computing [14], [15].

Based on the above procedures for the preparation of the safety features we have created a model for data security in the company. Security model was divided into 7 modules. For each module, we identify the main actors and modules are described by use cases. Applying knowledge and opportunities offered in the field of security data into the UML, we have created a new model for data security. Model we have compiled on the basis of possible situations that occur in the event of data security. Using this model in practice when implementing cloud can enhance the security of data stored in corporate cloud. In our study, we also offer new opportunities for solving the problem of security.

## AUTHORS: SHORT BIOGRAPHY

**1)  Mgr. Zuzana Priš áková,**
PhD Scholar, System engineering and informatics, Economic informatics, Department of
Informatics,    Mendel    University    in    Brno,    Czech    Republic.
Email:zuzana.priscakova@hotmail.com

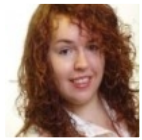

**2)  doc. Ing. Ivana Rábová, Ph.D.**
Academic staff – Associate Professor, Department of Informatics, Mendel University in
Brno, Czech Republic. Email:rabova@mendelu.cz

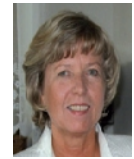